\documentclass[twocolumn,secnumarabic,amssymb, ,nofootinbib,nobibnotes, aps, prd]{revtex4}

\usepackage{graphicx,epstopdf,color}
\usepackage{amsmath,amsthm,amssymb}

\newcommand{\be}{\begin{equation}}
\newcommand{\ee}{\end{equation}}
\newcommand{\ba}{\begin{eqnarray}}
\newcommand{\ea}{\end{eqnarray}}

\def\ap{\approx}
\def\d{{\rm d}}

\def\lsim{\raise0.3ex\hbox{$\;<$\kern-0.75em\raise-1.1ex\hbox{$\sim\;$}}}
\def\gsim{\raise0.3ex\hbox{$\;>$\kern-0.75em\raise-1.1ex\hbox{$\sim\;$}}}
\def\eps{\varepsilon}
\def\theta{\vartheta}

\begin{document}

\title{Neutrino yield from Galactic cosmic rays}

\author{M.~Kachelrie\ss$^{1}$ and S.~Ostapchenko$^{2,3}$}

\affiliation{$^1$Institutt for fysikk, NTNU,  7491 Trondheim, Norway}
\affiliation{$^2$Hansen Experimental Physics Laboratory \&
Kavli Institute for Particle Astrophysics and Cosmology,
Stanford University, Stanford, CA 94305, U.S.A}
\affiliation{$^3$D.~V.~Skobeltsyn Institute of Nuclear Physics,
 Moscow State University, 119991 Moscow, Russia}

\begin{abstract}
We calculate the neutrino yield from collisions of cosmic ray (CR) nuclei
on gas using the event generator QGSJET-II. We present first the
general characteristics and numerical  results for the neutrino yield 
assuming power-law fluxes for the primary CR nuclei. Then we use 
three parameterisations for the Galactic CR flux to derive
the neutrino yield for energies around and above the knee.
The shape and the normalization of the resulting neutrino flux above 
$\sim 10^{14}$\,eV depends on the composition of the Galactic 
CR flux employed, but is generally dominated by its  proton component.
The spectral shape and magnitude of the neutrino flux suggest that the 
IceCube excess is not connected to interactions of Galactic sea CRs.
If a fraction of these events has a Galactic origin, then they may 
be caused by CR overdensities around recent close-by Galactic sources.
\end{abstract}

\date{May 15, 2014}


\maketitle


\vspace{3pc}


\section{Introduction}
\label{Introduction}

The IceCube Collaboration recently announced evidence for the first detection
of extraterrestrial neutrinos at the $4\sigma$ confidence 
level~\cite{Aartsen:2013pza}. This announcement followed the observation of 
two PeV neutrino cascades~\cite{Aartsen:2013bka}. The combined data set 
consists of 28 events with detected energy in the  range between 30\,TeV 
and 2\,PeV, while $10.6^{+5.0}_{-3.6}$ background events are  expected from 
atmospheric muons and neutrinos. This excess of events 
(denoted ``IceCube excess'' in the following) is 
consistent with a diffuse intensity (summed over flavors) at the level of
\begin{equation}\label{eq:ICnu}
E_\nu^2\, I_{\nu} \simeq (36\pm 12)\times
 {\rm eV}\,{\rm cm}^{-2}\,{\rm s}^{-1}\,{\rm sr}^{-1}\,,
\end{equation}
based on 17 events in the 60~TeV to 2~PeV energy range.

The implications of this ground-breaking discovery have been discussed 
widely~\cite{some,2.4,Ahlers:2013xia,Fox:2013oza,Neronov:2013lza,GZK2}. 
In particular, the non-observation of events beyond 
2\,PeV has been interpreted as a break or an exponential cutoff in the 
neutrino flux, but a single power-law with $I(E)\sim E^{-2.3}$ is compatible
with the data~\cite{2.4}.
Such a suppression, if present, is not typical for many models of extragalactic 
neutrino sources: For instance, if active galactic nuclei or $\gamma$-ray 
bursts were  sources of ultrahigh energy cosmic rays, then they
should produce neutrinos with energies up to $E\sim 10^{19}$\,eV.
The same holds for the cosmogenic neutrino flux which peaks around
$10^{18}$\,eV  for proton primaries~\cite{GZK2,GZK1}. Note also that the 
intensity (\ref{eq:ICnu}) practically saturates the cascade bound on
extragalactic neutrino intensity derived 
in Ref.~\cite{GZK1} from the Fermi-LAT observations of the diffuse
gamma-ray background\footnote{While the cascade bound is derived considering
CR interactions with the extragalactic $\gamma$-ray background during
propagation, a similiar bound applies to many neutrino sources themselves.}.

Alternatively, the neutrino events may have a Galactic origin.  The neutrino
excess has been associated e.g.\ with unidentified TeV $\gamma$-ray 
sources~\cite{Fox:2013oza}, CR pevatrons~\cite{Neronov:2013lza} or PeV dark 
matter~\cite{DM}.   
The Galactic neutrino flux contains a guaranteed component which is
produced by CRs interacting with gas during their confinement in a CR halo. 
This minimal Galactic neutrino flux has been discussed since
the late 1970s~\cite{70}.

The maximal energy of neutrinos produced via pion production in proton-photon 
($p\gamma$) or proton-gas  ($pp$) interactions is approximately 10\% of the 
energy of the proton primary. 
Thus the observed 2\,PeV neutrino event requires proton 
energies above 20\,PeV.  However,  already at $10^{15}$\,eV  protons  
represent only a subdominant fraction of the primary  CR flux compared to 
helium and heavier nuclei~\cite{Antoni:2005wq,Ahn:2010gv}.  Moreover, the 
composition of the CR flux becomes increasingly heavier in the energy range 
between the knee at $E_{\rm k} \ap 4$\,PeV and 
$10^{17}$\,eV~\cite{Antoni:2005wq,IceCube:2012vv,Apel:2013dga}. 
Since the maximal neutrino energy in nucleus-proton collisions is a factor
$A$ lower than in $pp$ interactions ($A$ being the nuclear mass number), 
 the required minimal CR energy to explain
the IceCube events increases compared to $pp$ processes. This implies in turn 
that the number of potential scattering events, and thus secondary fluxes, is
drastically reduced, because the CR spectrum is steeply falling,  
$I(E)\propto E^{-3.1}$ above the  knee~\cite{Antoni:2005wq}. 
Therefore it is essential to account correctly for
the elemental composition of the Galactic CR flux, if one aims at
relating the neutrino intensity required to explain the IceCube excess to 
the primary CR intensity~\cite{Candia:2003ay}.

Aim of this work is to quantify the neutrino yield from nucleus-proton
collisions  using up-to-date simulation tools for the relevant
hadron production processes and including information
on the elemental composition of the CR flux around and above the knee
region. 
Our simulations are based on the event generator QGSJET-II-04~\cite{qgs}
which includes relevant experimental information from run I of 
LHC~\cite{ost12}. We present in Sec.~II the general characteristics of 
the neutrino yields for the case of  power-law fluxes of primary nuclei.
Then we calculate in Sec.~III the neutrino yields for three parametrisations 
of the CR flux, suitable for the energy region of the knee. Finally, we 
comment in Sec.~IV on the IceCube neutrino excess in view of our
findings before we conclude.

\section{Neutrino yield for power-law CR fluxes}

The neutrino intensity $I_\nu(E)$ produced by CR interactions with the 
interstellar gas is given by
\ba
I_\nu(E) =\tilde\eps_{\rm M}\: d_{\rm CR}\,n_{\rm gas}\: 
\sum _i \int_{A_iE}^{\infty}dE' \; I_i(E')\:
\sigma ^{\rm inel}_{A_ip}(E'/A_i)  && \nonumber \\
              \times       \frac{dn_{A_ip\rightarrow \nu}(E'/A_i,E)}
		     {dE} \,,&&\label{phi-nu}
\ea
where the sum goes over the primary CR mass groups $i$,  $I_i(E)$ is the 
partial intensity for the $i$-th   group,  $n_{\rm gas}$ is the 
gas density, and $d_{\rm CR}$ is the path-length travelled by the CRs. For
simplicity, we assume here a uniform gas density and neglect
the dependence of the confinement time on the charge of the nuclei.  
We account for the helium contribution in the interstellar medium by 
means of an enhancement factor\footnote{Note that in contrast to its
usual definition, see e.g.~\cite{KOM},  $\tilde\eps_{\rm M}$ 
accounts only for nuclei in the interstellar medium.}, 
$\tilde\eps_{\rm M}\simeq 1.3$ around $E_\nu\sim {\rm PeV}$,
as explained in the following. Particle physics enters via 
the inelastic cross section $\sigma^{\rm inel}_{Ap}(E_A)$ for an
interaction of a nucleus of mass number $A$ and energy per nucleon $E_A$
with a proton, and the neutrino production spectrum 
$d n_{Ap\rightarrow \nu}(E_A,E)/d E$ per inelastic event. The latter is defined
as the convolution of the production spectra for different hadron species
and the spectra for their decays into neutrinos,
\ba
 \frac{d n_{Ap\rightarrow \nu}(E_A,E)}{d E}
 =\sum _h\int dE_h\; \frac{d n_{Ap\rightarrow h}(E_A,E_h)}{d E_h}  && \nonumber \\
 \times \frac{d n^{\rm dec}_{h\rightarrow \nu}(E_h,E)}{d E}\, .&&\label{conv}
\ea
Introducing the energy fraction $z= E/(E'/A_i)$ of the produced neutrinos in
Eq.~(\ref{phi-nu}), we can rewrite it in the case of power-law energy 
spectra of the CRs, $I_i(E)\propto E^{-\alpha_i}$, as
\be 
  I_\nu(E) =\tilde\eps_{\rm M}\: d_{\rm CR}\,n_{\rm gas} \sum _i I_i(E)\:
  Z_{A_i}^{\nu}(E,\alpha _i)\,, \label{phi-nu-z}
\ee
where the so-called 
$Z$-factors\footnote{Originally, $Z$-moments were introduced 
calculating inclusive fluxes of high energy muons and neutrinos
resulting from CR interactions in the atmosphere,   without the factor
$\sigma^{\rm inel}_{Ap}$.}~\cite{gaisser-book}
$Z_A^{\nu}$ for neutrino production are defined as
\ba
Z_{A}^{\nu}(E,\alpha) =A^{-\alpha}\:\int _0^1 dz\;z^{\alpha -1}\:
 \sigma ^{\rm inel}_{Ap}(E/z) && \nonumber \\
 \times \frac{d n_{Ap\rightarrow \nu}(E/z,z)}{d z}\, .&&\label{zfac-nu}
\ea 
Using (\ref{conv}), it is convenient to express $Z_{A}^{\nu}$ via the
$Z$-factors for hadron production as follows
\be
Z_{A}^{\nu}(E,\alpha) =\sum _h \int _0^1 dz_{\nu}\;z_{\nu}^{\alpha -1}\:
f^{\rm dec}_{h\rightarrow \nu}(z_{\nu}) \; Z_{A}^{h}(E/z_{\nu},\alpha)\,
,\label{zfac-dec}
\ee
where $Z_{A}^{h}$ are defined similarly to $Z_{A}^{\nu}$ as
\ba
Z_{A}^{h}(E,\alpha) =A^{-\alpha}\:\int _0^1 dz\;z^{\alpha -1}\:
 \sigma ^{\rm inel}_{Ap}(E/z) && \nonumber \\
 \times \frac{d n_{Ap\rightarrow h}(E/z,z)}{d z}\, &&\label{zfac-h}
\ea 
and $f^{\rm dec}_{h\rightarrow \nu}$ are the (Lorentz-invariant)
spectra for hadron decays into neutrinos,
\be
 f^{\rm dec}_{h\rightarrow \nu}(z_{\nu}^+) =
 \frac{d n^{\rm dec}_{h\rightarrow \nu}(E,z_{\nu}^+)}{d z_{\nu}^+} \,.
\ee
In deriving (\ref{zfac-dec}), we used the high-energy limit 
\be
 z_{\nu}^+\equiv (E_{\nu}+p_{z_{\nu}})/(E_{h}+p_{z_{h}})\simeq
 E_{\nu}/E_{h}=z_{\nu}\, .
\ee
Thus, all the dependence on the properties of proton-proton and nucleus-proton
interactions is contained in the hadronic  $Z$-factors $Z_A^{h}$. They also
define how strongly the contribution to neutrino production from primary CR 
nuclei is suppressed relative to the one of protons. Due to the steep
slopes $\alpha _i$ of the primary   spectra, these  $Z$-factors are dominated
by the hadron spectra in the very forward direction. This allows one to
estimate the  suppression of the nuclear contribution,
using the well-known relation for the mean number of interacting 
(``wounded'') projectile nucleons
 $\langle n_{AB}^{{\rm w}_p}\rangle$ for nucleus $A$ -- nucleus $B$ 
collisions~\cite{Bialas1976},
\be
\langle n_{AB}^{{\rm w}_p}(E)\rangle
=\frac{A\:\sigma_{pB}^{\rm  inel}(E)}{\sigma_{AB}^{\rm  inel}(E)}\, .
\label{nwounded}
\ee
This relation holds both in the Glauber approach and in Reggeon Field 
Theory, if one neglects the contribution of target 
diffraction~\cite{Kalmykov1993}. 

Thus, for the forward ($z\rightarrow 1$) spectra of secondary hadrons we get
\ba
 \frac{d n_{Ap\rightarrow h}(E,z)}{d z}\simeq \langle n_{Ap}^{{\rm
 w}_p}(E)\rangle \;
  \frac{d n_{pp\rightarrow h}(E,z)}{d z} && \nonumber \\
  =A\;\frac{\sigma_{pp}^{\rm  inel}(E)}{\sigma_{Ap}^{\rm  inel}(E)}\;
   \frac{d n_{pp\rightarrow h}(E,z)}{d z}\, .\label{forw-spec}
 \ea
 Substituting (\ref{forw-spec}) into (\ref{zfac-h}), we obtain
\be
Z_{A}^{h}(E,\alpha)\simeq  A^{1-\alpha}\;Z_{p}^{h}(E,\alpha)\, ,
\label{zfac-superp}
\ee
which leads in turn to 
\be 
  I_\nu(E) \simeq \tilde\eps_{\rm M}\: d_{\rm CR}\,n_{\rm gas} \sum _i I_i(E)\:
  Z_{p}^{\nu}(E,\alpha _i)\;  A_i^{1-\alpha_i}\,. \label{phi-nu-end}
\ee
Thus the simple $A_i^{1-\alpha_i}$ rule which is often applied to convert
neutrino fluxes from $pp$ collisions into those of $Ap$ collisions is only
modified by the ratio of  $Z$-factors  
$Z_{p}^{\nu}(E,\alpha _i)/Z_{p}^{\nu}(E,\alpha _p)$, if the
spectral slopes for CR nuclei differ from the one for protons,
$\alpha_i \neq \alpha_p$.

\begin{table*}[t]
\begin{tabular*}{1\textwidth}{@{\extracolsep{\fill}}l|cccc|cccc}
\hline 
Primary     
 &\multicolumn{4}{c|}{$A^{\alpha -1}\;Z_{A}^{\pi ^{\pm}}\!(E,\alpha)$}  
 &\multicolumn{4}{c}{$A^{\alpha -1}\;Z_{A}^{K ^{\pm}}\!(E,\alpha)$}  
  \\
 nucleus&    $10^{5.5}$ GeV &  $10^{6}$ GeV &  $10^{6.5}$ GeV & $10^{7}$ GeV & 
 $10^{5.5}$ GeV &  $10^{6}$ GeV &  $10^{6.5}$ GeV & $10^{7}$ GeV
  \tabularnewline
\hline 
\hline 
 p   &  
 1.6   & 1.8   & 1.8    & 1.9   & 0.23    &  0.24    & 0.28    &  0.29  \tabularnewline
$^4$He & 
 1.7   & 1.8   & 1.9    & 2.0   & 0.23    &  0.25    & 0.28    &  0.29  \tabularnewline
 $^{14}$N &  
 1.7   & 1.8   & 1.9    & 2.0   & 0.23    &  0.25    & 0.28    &  0.30  \tabularnewline
 $^{25}$Mn & 
 1.7   & 1.9   & 1.9    & 2.0   & 0.23    &  0.25    & 0.28    &  0.30  \tabularnewline
$^{56}$Fe &  
 1.7   & 1.9   & 1.9    & 2.0   & 0.23    &  0.26    & 0.28    &  0.30  \tabularnewline
\hline 
\end{tabular*}
\caption{$Z$-factors $Z_{A}^{h}(E,\alpha)$ for charged
pion ($h=\pi^{\pm}$) and kaon ($h=K^{\pm}$) production,
  multiplied by $A^{\alpha -1}$, 
  as a function of energy $E$, as calculated using
   QGSJET-II-04  for different primary nuclei, for
    $\alpha =3.1$.\label{tab: z-factor-qgs}}
\end{table*}
\begin{table*}[tb]
\begin{tabular*}{1\textwidth}{@{\extracolsep{\fill}}l|ccccc|ccccc}
\hline 
Primary     
 &\multicolumn{5}{c|}{$A^{\alpha -1}\;Z_{A}^{\pi ^{\pm}}\!(E,\alpha)$}  
 &\multicolumn{5}{c}{$A^{\alpha -1}\;Z_{A}^{K ^{\pm}}\!(E,\alpha)$}  
  \\
 nucleus&    $\alpha=2$  &  $\alpha=2.5$ & $\alpha=3$ & $\alpha=3.5$ & $\alpha=4$ & 
 $\alpha=2$  &  $\alpha=2.5$ & $\alpha=3$ & $\alpha=3.5$ & $\alpha=4$
  \tabularnewline
\hline 
\hline 
p   &  
 22   & 5.7   & 2.1    & 0.98   & 0.51  & 3.1    &  0.82    & 0.29    &  0.13   &  0.064 \tabularnewline
$^4$He & 
 22   & 5.8   & 2.2    & 0.99   & 0.51  & 3.1    &  0.83    & 0.30    &  0.13   &  0.065 \tabularnewline
 $^{14}$N &  
 22   & 5.8   & 2.2    & 0.99   & 0.51  & 3.1    &  0.83    & 0.30    &  0.13   &  0.065 \tabularnewline
 $^{25}$Mn & 
 22   & 5.8   & 2.2    & 1.0    & 0.54  & 3.1    &  0.85    & 0.30    &  0.13   &  0.066 \tabularnewline
$^{56}$Fe &  
 22   & 5.8   & 2.2    & 1.0    & 0.54  & 3.1    &  0.85    & 0.30    &  0.14   &  0.068 \tabularnewline
\hline 
\end{tabular*}
\caption{Same as in  Table~\ref{tab: z-factor-qgs} for different slopes $\alpha$ of the primary spectra
for $E=10^6$ GeV.\label{tab: z-factor-qgs-alpha}}
\end{table*}

In Table~\ref{tab: z-factor-qgs}, we present the $Z$-factors 
$Z_{A}^{h}(E,\alpha)$ multiplied by $A^{\alpha -1}$ for charged
pion and kaon production\footnote{For the $Z$-factors of kaons, the
relation $Z_{A}^{K_{\rm L}}\simeq Z_{A}^{K^{\pm}}/2$ holds.}
by different primary nuclei,  calculated with the QGSJET-II-04 model
for the primary CR slope $\alpha =3.1$.
Additionally, we demonstrate in Table~\ref{tab: z-factor-qgs-alpha}
the dependence of these factors
  on the  slopes $\alpha$ of the primary spectra for $E=10^{6}$ GeV.
    As one can see from the Tables,  Eq.~(\ref{zfac-superp}) holds 
here to a  very good accuracy:
The    factors $A^{\alpha -1}\,Z_{A}^{h}(E,\alpha)$ 
for different primary particles 
 agree with each other to better than 10\% accuracy. The relatively weak
 energy-dependence of these factors stems from the energy rise of the 
 inelastic nucleus-proton cross sections.

\begin{table*}[t]
\begin{tabular*}{1\textwidth}{@{\extracolsep{\fill}}l|ccccc|ccccc}
\hline 
Primary     
 &\multicolumn{5}{c|}{$h=\pi ^{\pm}$}   &\multicolumn{5}{c}{$h=K ^{\pm}$}  
  \\
 nucleus&    $\alpha=2$  &  $\alpha=2.5$ & $\alpha=3$ & $\alpha=3.5$ & $\alpha=4$ & 
 $\alpha=2$  &  $\alpha=2.5$ & $\alpha=3$ & $\alpha=3.5$ & $\alpha=4$
  \tabularnewline
\hline 
\hline 
p   &  
3.1   & 2.9   & 2.7    & 2.6   & 2.5  & 3.2    &  3.0    & 2.9   &  2.8   &  2.8 \tabularnewline
$^4$He & 
2.9   & 2.7   & 2.6    & 2.5   & 2.5  & 2.9    &  2.7     & 2.7   &  2.7   &  2.7 \tabularnewline
\hline 
\end{tabular*}
\caption{Ratio of  $Z$-factors
 $Z_{A|{\rm He}}^{h}\!(E,\alpha)/Z_{A}^{h}\!(E,\alpha)$ 
for hadron $h$ production  on helium
 and proton targets  for different slopes $\alpha$ of the primary spectra;
 $E=10^6$ GeV.\label{tab: z-ratios}}
\end{table*}

We also use the ratios of $Z$-factors for hadron production on helium
and proton targets  $Z_{A|{\rm He}}^{h}/Z_{A}^{h}$, as compiled in 
Table \ref{tab: z-ratios} for $E=10^6$ GeV, to determine the 
enhancement factor $\tilde\eps_{\rm M}$ which accounts for the contribution
to neutrino production from CR interactions with
the helium component of the ISM.
Here $Z_{A|{\rm He}}^{h}$ is defined by Eq.\  (\ref{zfac-h})
with the replacements 
$\sigma ^{\rm inel}_{Ap} \rightarrow \sigma ^{\rm inel}_{A{\rm He}}$
and $d n_{Ap\rightarrow h}/dz \rightarrow d n_{A{\rm He}\rightarrow h}/dz$.
For $\alpha \simeq 2.5-3$, we have
$Z_{A|{\rm He}}^{h}/Z_{A}^{h}\simeq 2.7-3$ for all CR primaries;
typical deviations do not exceed the 10\% level\footnote{Since both 
$Z_{A|{\rm He}}^{h}$ and $Z_{A}^{h}$ are defined for the same spectra of
primary mass groups, the effects of primary abundances and of the 
energy-dependence of these $Z$-factors are largely cancelled in their ratios
in the expression (\ref{enh-factor}) for the enhancement factor
 $\tilde\eps_{\rm M}$.}  [c.f.\ Eq.~(\ref{zfac-superp})
and  Table~\ref{tab: z-ratios}]. For the He/H abundance ratio in the ISM
$R_{\rm He/H}\simeq 0.096$ \cite{Meyer1985}, we thus obtain
 \be
 \tilde\eps_{\rm M}=1 + R_{\rm He/H}\,\frac{Z_{A|{\rm He}}^{h}}{Z_{A}^{h}}\simeq
 1.3\,.\label{enh-factor}
 \ee
 
 To check the model dependence of our results,
we repeated the same calculations as in Table~\ref{tab: z-factor-qgs}
for $E=10^{5.5}$ and $10^7$ GeV, using the EPOS-LHC
model~\cite{werner06,pierog13}, the results being collected in 
Table~\ref{tab: z-factor-epos}.
\begin{table}[tbh]
\begin{tabular*}{0.48\textwidth}{@{\extracolsep{\fill}}l|cc|cc}
\hline
Primary     
 &\multicolumn{2}{c|}{$A^{\alpha -1}\;Z_{A}^{\pi ^{\pm}}\!(E,\alpha)$}  
 &\multicolumn{2}{c}{$A^{\alpha -1}\;Z_{A}^{K ^{\pm}}\!(E,\alpha)$}  
  \\
 nucleus&     $10^{5.5}$ GeV & $10^{7}$ GeV &  $10^{5.5}$ GeV &  $10^{7}$ GeV
  \tabularnewline
\hline 
\hline 
p   &       1.8   & 2.3     & 0.27    &  0.36  \tabularnewline
$^4$He &      1.9   & 2.4     & 0.27    &  0.36  \tabularnewline
 $^{14}$N &   2.0   & 2.7     & 0.27    &  0.36  \tabularnewline
 $^{25}$Mn &  2.1   & 2.7     & 0.27    &  0.37  \tabularnewline
$^{56}$Fe &   2.2   & 2.7     & 0.27    &  0.37  \tabularnewline
\hline 
\end{tabular*}
\caption{Same as in Table~\ref{tab: z-factor-qgs} for the EPOS-LHC model
and for two values of energy $E$.\label{tab: z-factor-epos}}
\end{table}
For EPOS-LHC, the deviations from  Eq.\ (\ref{zfac-superp}) are
more significant, reaching 20\% in the case of primary iron.
More importantly, also the predicitions 
for the $Z$-factors $Z_{p}^{h}$  for primary protons
 by EPOS-LHC are $\simeq 20\%$ higher than by  QGSJET-II-04.
Thus this range defines the characteristic uncertainty of our results
that arises from the treatment of hadronic interactions.

\section{Neutrino flux in the knee region}

The elemental composition of the CR flux below $E\sim 10^{14}$\,eV is
relatively well determined and can be
be described to first order by power-laws.
At higher energies, the low CR flux prevents direct measurements, 
and the elemental composition of the CR flux becomes rather uncertain; 
for a review of experimental methods and results see e.g.\
Ref.~\cite{Bluemer:2009zf}. While there exist yet substantial uncertainties 
concerning the partial contributions of different mass groups to the 
primary CR  composition, there is a general agreement that the knee in 
the total CR spectrum at 
$E_{\rm k} \ap 4$\,PeV coincides with a suppression of the primary proton 
flux, and that the composition becomes increasingly heavier in the energy 
range between the knee  and 
$10^{17}$\,eV~\cite{Antoni:2005wq,Apel:2013dga,IceCube:2012vv,Aglietta:2004}.

Explanations for the origin of the knee fall in three main categories. 
First, there have been speculations that
interactions may change in the multi-TeV region and the CR flux
may be suppressed because of additional energy loss channels. This
possibility is now excluded by LHC data~\cite{d'Enterria:2011kw}. 
Second, the knee may correspond 
to the maximum rigidity to which CRs can be accelerated by the dominant
population of Galactic CR sources~\cite{pop,Hillas05}. 
Third, the knee energy may correspond to the rigidity at which the CR Larmor 
radius $r_{\rm L}$ is of the order of the coherence length $l_{\rm c}$ of the 
turbulent magnetic field in the Galactic disk. As a result, a transition 
from large-angle to small-angle scattering or Hall diffusion is expected, 
the energy dependence of the confinement time changes which in turn induces 
a steepening of the CR 
spectrum~\cite{Ptuskin1993,Candia:2002we,Candia:2003dk,GKS14}.

Both the second and third possibilities lead to a rigidity-dependent 
sequence of knees at $ZE_{\rm k}$, a behavior first suggested by 
Peters~\cite{Peters}.
 In contrast to models in category 2, those of category 
3 predict both the position of the knee and the rigidity dependent 
suppression of the different CR components for a given model of the 
Galactic magnetic field~\cite{GKS14}.


Various models which describe the elemental composition of the total CR 
flux have been developed.  The parametrization of Honda and 
Gaisser~\cite{HG02}  is widely 
used in calculations of the atmospheric neutrino flux up to energies 
$\sim 100$\,GeV. Since it does not attempt to include the CR knee, 
it cannot be extrapolated into the energy range of our interest. 
The poly-gonato model~\cite{Hoerandel:2002yg} is a fit of 
rigidity-dependent knees at $E=ZE_{\rm k}$ to measurements of the 
total CR intensity. Below and above $E=ZE_{\rm k}$, the fluxes of 
individual CR nuclei are assumed to follow
power-laws $\phi_A(E)=K_{A,1} E^{-\gamma_{A,1}}$ and 
$\phi_A(E)=K_{A,2} E^{-\gamma_{A,2}}$ which are smoothly interpolated.
The fluxes are assumed to  steepen by a common amount, 
$\gamma_{A,2}=\gamma_{A,1}+\delta$ with $\delta\simeq 2.1$, for more details 
see~\cite{Hoerandel:2002yg}.

\begin{figure}
\begin{center}
\includegraphics[width=0.95\columnwidth,angle=0]{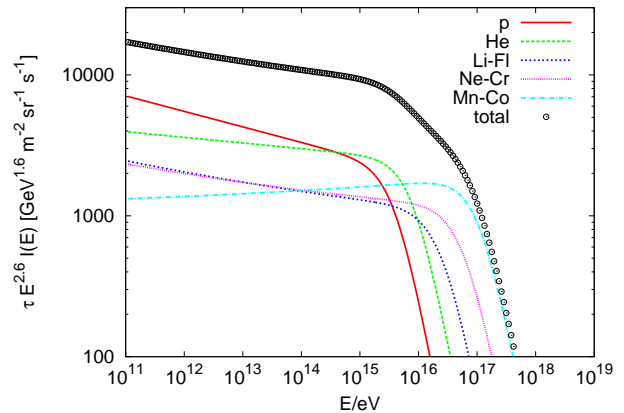}
\end{center}
\caption{All particle CR spectrum $E^{2.6}I(E)$ and individual 
contributions of five elemental groups in the poly-gonato model
as a function of the primary energy $E$.
\label{fig:CRpoly}}
\end{figure}

\begin{figure}
\begin{center}
\includegraphics[width=0.95\columnwidth,angle=0]{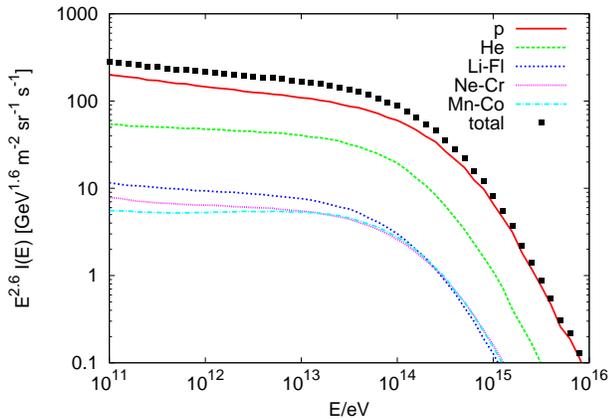}
\end{center}
\caption{Total neutrino spectrum $E^{2.6}I(E)$ and individual contributions 
of five elemental groups for $X=30\,$g/cm$^2$ in the poly-gonato 
model.
\label{fig:nupoly}}
\end{figure}

The CR intensity in this model, split into five elemental groups, is shown in 
Fig.~\ref{fig:CRpoly}. We neglected elements with $A>56$ which give in
the poly-gonato model~\cite{Hoerandel:2002yg} an important contribution
to the Galactic CR intensity at the highest energies, because their 
contribution 
to the neutrino intensity is---as expected from  Eq.~(\ref{zfac-superp})
and as we will see in the following---negligible.
The steepening of the CR intensity around the knee 
is very pronounced in this model, $\delta\sim 2$, and as a result 
the composition is heavier than suggested by KASCADE-Grande data.  
The resulting neutrino intensity  $I_\nu(E)$ is shown in Fig.~\ref{fig:nupoly},
where we assumed that the CR nuclei cross the grammage $X=30\,$g/cm$^2$; this
value corresponds for primary protons with energy 1\,PeV 
to the interaction depth $\tau_{pp}=1$. 
Above few hundred  TeV, the intensity is  dominated by the proton contribution
and is strongly decreasing with energy as  $I_\nu(E)\sim E^{-4.7}$.
Therefore the expected neutrino energy distribution disagrees with the 
neutrino spectrum suggested by the IceCube excess, in particular with
the two PeV events.

The Hillas model~\cite{Hillas05}  and its variants belong to the category~2, 
associating
the knee with the maximal rigidity achievable in the dominant population
of Galactic CR sources. Moreover, the Hillas model assumes that the ankle
signals the transition from Galactic to extragalactic CRs. Therefore,
an additional population of Galactic CR sources must exist (``the component
B'' of Ref.~\cite{Hillas05}) which fills the gap between the knee and the 
ankle. Thus the Hillas model contains two Galactic components.
Each population is assumed to contain five elemental groups and cuts off
at a characteristic rigidity.

\begin{figure}
\begin{center}
\includegraphics[width=0.95\columnwidth,angle=0]{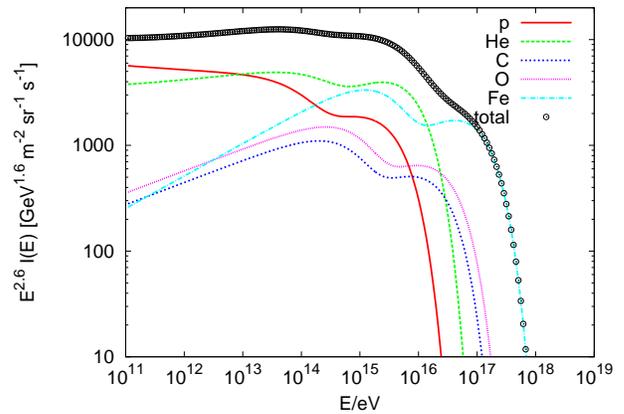}
\end{center}
\caption{All particle CR spectrum  $E^{2.6}I(E)$ and individual 
contributions of five elemental groups in the Hillas model
as a function of the primary energy.
\label{fig:CRHillas3}}
\end{figure}

\begin{figure}
\begin{center}
\includegraphics[width=0.95\columnwidth,angle=0]{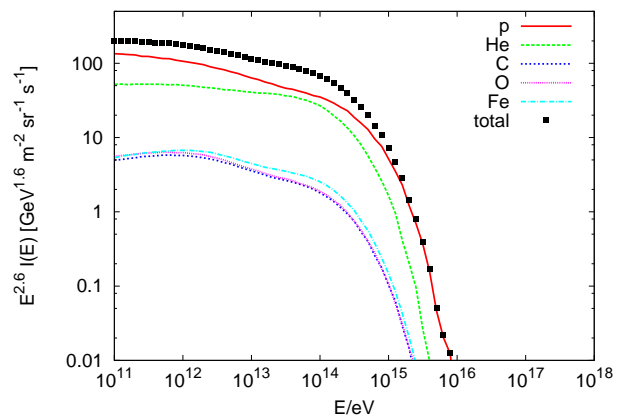}
\end{center}
\caption{Total neutrino spectrum $E^{2.6}I(E)$ and individual contributions 
of five elemental groups  for $X=30\,$g/cm$^2$ in the Hillas model.
\label{fig:nuHillas3}}
\end{figure}

Variations of the original Hillas model were presented in Refs.~\cite{Ga12,GST}.
We use here the new parametrisation  H3a given in Tab.~3 of Ref.~\cite{GST}, 
where the first population is fitted only above $E/Z=200$\,GeV, i.e.\ 
above the hardening observed by CREAM~\cite{Ahn:2010gv} and 
PAMELA~\cite{Adriani:2011cu}. The intensity of Galactic CRs in this model 
is shown in Fig.~\ref{fig:CRHillas3}, the
resulting neutrino intensity  in Fig.~\ref{fig:nuHillas3}.
The neutrino intensity is again dominated by the proton contribution;
its shape is very similar to the neutrino intensity of the polygonato
model.

\begin{figure}
\begin{center}
\includegraphics[width=0.95\columnwidth,angle=0]{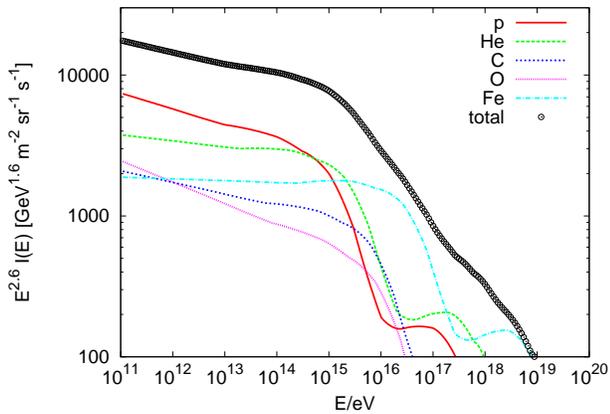}
\end{center}
\caption{All particle CR spectrum and individual contribution of
five elemental groups  in the escape model of Ref.~\cite{GKS14}.
\label{fig:CRGKSb}}
\end{figure}

\begin{figure}
\begin{center}
\includegraphics[width=0.95\columnwidth,angle=0]{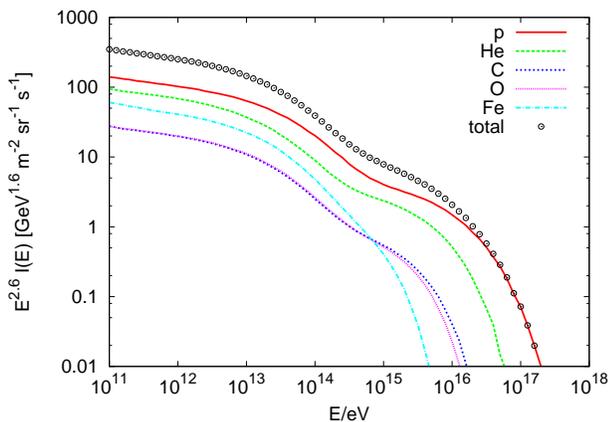}
\end{center}
\caption{Total neutrino spectrum $E^{2.6}I(E)$ and individual contributions 
of five elemental groups for $X=30\,$g/cm$^2$ in the escape model of 
Ref.~\cite{GKS14}.
\label{fig:nuGKSb}}
\end{figure}

Finally, we consider a parametrisation of the CR flux motivated by
the recent results of Ref.~\cite{GKS14}: There, the escape of CRs 
from our Galaxy was studied calculating trajectories of individual CRs
in models of the regular and turbulent Galactic magnetic field. For a 
coherence length $l_{\rm c} \simeq (2-5)$\,pc of the turbulent field and a 
reduced turbulent magnetic field, a knee-like structure at 
$E/Z={\rm few}\times 10^{15}$\,eV was found,
which is sufficiently strong to explain the proton knee observed by KASCADE.
The resulting intensity of four other elemental groups are shown
in Fig.~\ref{fig:CRGKSb}. They are
consistent with the energy spectra of CR nuclei determined by KASCADE 
and KASCADE-Grande. 
The resulting neutrino intensity for $X=30\,$g/cm$^2$ is shown in 
Fig.~\ref{fig:nuGKSb}. The suppression of the neutrino intensity
above the knee is less pronounced as in the previous models,
since the decrease of the CR escape time $\tau_{\rm esc}(E)$ slows 
down around $E/Z\simeq 10^{16}$\,eV for a weak turbulent field.
In the energy range between $10^{13}$\,eV and  $10^{16}$\,eV, the
neutrino intensity scales as $I_\nu(E)\propto E^{-3.2}$.

\begin{figure}
\begin{center}
\includegraphics[width=0.95\columnwidth,angle=0]{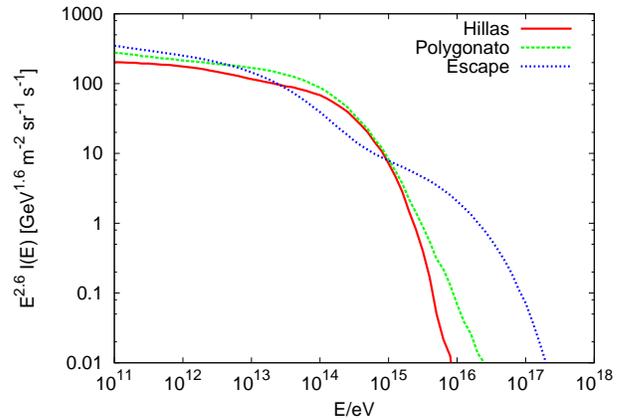}
\end{center}
\caption{A comparison of the total neutrino spectra $E^{2.6}I(E)$ 
predicted by the three CR models.\label{fig:nutot}}
\end{figure}

In summary, we found that the neutrino intensity below $10^{14}$\,eV
reflects the slope of protons and agrees therefore in all three
models. In contrast, the exact position of  the ``neutrino knee''
 and the slope  of
neutrino intensity above this break depends on the nuclear composition
and is therefore model dependent: A comparison of 
the neutrino intensity in the three models is shown in Fig.~\ref{fig:nutot}.

Finally, one may yet ask which of the three composition models considered
is the more realistic one in the light of recent CR data. 
Comparing e.g.\ the CR spectra predicted by
 the Polygonato model to the intensities of individual groups of CR nuclei 
 up to $10^{17}$\,eV, measured by the KASCADE and KASCADE-Grande
 experiments~\cite{Antoni:2005wq,Apel:2013dga}, already an inspection by eye 
indicates that this model predicts a too heavy composition above the knee.
The Hillas model of Ref.~\cite{GST} describes well
the average composition (represented e.g.\ by $\ln(A)$) but fails to
reproduce the up-turn of the light component around $10^{17}$ eV
observed by KASCADE-Grande~\cite{Apel:2013dga,KG-prd13}.
By contrast, such an up-turn around $E/Z\simeq 10^{16}$\,eV is the 
characteristique feature of the escape model~\cite{GKS14}. As a result, 
the intensity of individual groups of CR nuclei measured by
KASCADE and KASCADE-Grande~\cite{Antoni:2005wq,Apel:2013dga} is well
reproduced in this model~\cite{GKS14}.

\section{Comments on the IceCube events}

We discuss now briefly how our results impact the Galactic interpretation
of the IceCube excess.

\subsection{Neutrinos from Galactic Sea CRs}

The results of the previous section can be converted to the neutrino
intensity resulting from collisions of Galactic sea CRs on gas 
after evaluating the interaction depth
\be \label{tau}
 \tau_\nu = \sigma_{\rm inel}^{pp}(E) \;  
 \int_{\rm l.o.s.} \!\!\!\!\!\! \d s\, n(\rho,z) \,,
\ee
where $\rho=(x^2+y^2)^{1/2}$ is the distance from the Galactic center (GC) 
in the Galactic plane, $z$ the distance above the plane, and $s$ the 
distance from the Sun along the chosen line-of-sight (l.o.s.). We neglect
for our estimations the contribution from primary nuclei, because we have 
seen that the neutrino flux is dominated by $pp$  interactions.

The gas distribution in the Galactic disk can be modeled as 
$n(z)=N\exp(-(z/z_{1/2})^2)$ with $N\simeq 0.3$\,cm$^{-3}$ at $R_\odot$
(increasing to $N\simeq 10$\,cm$^{-3}$ at the GC) and $z_{1/2}= 0.21$\,kpc.
Integrating (\ref{tau}) with $\sigma_{\rm inel}^{pp}(E)\simeq 60$\,mb
results in the maximal interaction depth of $\tau\sim 0.005$ 
towards the GC.

At the reference energy $E_\ast=1$\,PeV, all three parametrisations
predict a neutrino intensity around 
$\tau_\nu E_\ast^{2.6}\,I_\nu(E_\ast)\sim 10$\,GeV$^{1.6}$ 
m$^{-2}$ s$^{-1}$  sr$^{-1}$, corresponding to
$E_\nu^2\, I_{\nu} \sim 0.1 \times
 {\rm eV}\,{\rm cm}^{-2}\,{\rm s}^{-1}\,{\rm sr}^{-1}$. Thus even in
the direction of the largest expected intensity, the predicted
neutrino intensity due to diffuse Galactic CR interactions
is about two orders of magnitude too small compared to the
IceCube excess. 
Moreover, the neutrino events should be concentrated within
$|b| \leq 1^\circ$~\cite{b}, reflecting the very slim Galactic plane, 
which is much narrower than the latitude distribution of the IceCube 
events.

\subsection{Neutrinos from Galactic CR sources}

We consider next the neutrino flux produced close to recent CR sources.
The propagation of CRs on distances $l> {\rm few}\times l_{\rm coh}$ can 
be approximated by 
diffusion~\cite{fil}.  For $l_{\rm coh}\sim 10$\,pc as found in 
Ref.~\cite{Iacobelli:2013fqa} for the Galactic disk, the diffusion approach
is marginally justified for the time scales, $10^3$ to $10^4$\,yr, we 
consider. 

Galactic accelerators able to produce CRs with energies $10^{16}$\,eV  
have typically only short life-times: For instance, the highest energy 
particles produced by a supernova remnant (SNR) are thought to escape at the
end of the Sedov phase after few 100\,yrs. Approximating therefore the
accelerator as a bursting source, the number density 
$n_{\rm CR}(E,r)=dN/(dE\,dV)$  of CRs at the distance $r$ is  given by
\begin{equation}
\label{eq:nburst}
 n_{\rm CR}(E,r)= \frac{Q(E)}{\pi^{3/2}\, r_{\rm diff}^3} \,
 \exp\left[-r^2/r_{\rm diff}^2 \right] ,
\end{equation}
with $r_{\rm diff}^2=4D\,t$ assuming no energy losses.  We use as diffusion 
coefficient $D(E_\ast)=3\times 10^{28}$\,cm$^2$/s at our reference energy 
$E_\ast=1$\,PeV and assume that the source injects instantaneously 
CRs with the total energy $E_{\rm p}=10^{50}$\,erg in protons with
an injection spectrum $Q(E)=Q_0 (E/E_0)^{-\alpha}$ 
between the minimal energy $E_0=1$\,GeV and a maximal energy 
$E_{\max}=10$\,PeV.  We choose $\alpha=2.0$ suggested by shock acceleration. 
Then PeV CRs are concentrated within 
$r_{\rm diff}\sim 35$\,pc after 3000\,yr. 

The brightest spots in the Galactic neutrino sky are likely giant molecular 
clouds (GMC) immersed into the CR overdensities close to recent CR sources.  
The neutrino flux from a point source at the distance $d$ is given by
\begin{equation}
 \phi_\nu(E)= \tilde\eps_{\rm M}
 \frac{c\,\sigma_{\rm inel}}{4\pi d^2} \frac{M_{\rm cl}}{m_p} \,
 n_{\rm CR}(E)\, Y_\nu(E) \,,
\end{equation}
where $Y_\nu(E)$ denotes the neutrino yield $Y_\nu(E)= \phi_\nu(E)/[\tau(E)\,\phi(E)]$. Assuming as cloud mass $M_{\rm cl}=10^5M_\odot$ and as 
distance $d=1$\,kpc results in the neutrino flux
\begin{equation}
 E^2 \phi_\nu(E) \simeq 140 \;
 {\rm eV}\,{\rm cm}^{-2}\,{\rm sr}^{-1}\,.
\end{equation}
Sources of this kind would be clearly visible on the neutrino sky as
seen by IceCube. Choosing as source rate $\dot N\simeq 1/(30\,{\rm yr})$,
which coincides with the Galactic SN rate, implies that the average number 
$N_s$ of such sources 
present in the Galaxy equals $N_s\simeq 100$. Using as volume of the
Galactic disk $V=\pi R^2 h \simeq 140$\,kpc$^3$, there are on average
0.5~sources within one kpc distance to an observer. We note also that
the presence of GMCs close to SNRs is not unnatural, since they are born most
likely in OB associations.

Any source of high-energy neutrinos produces also $\gamma$-rays. 
In Ref.~\cite{Neronov:2013lza},  the 1--10\,TeV $\gamma$-ray flux from known 
sources in the direction towards the GC was compared with the IceCube excess. 
Extrapolating the  $\gamma$-ray flux of these sources to higher energies implies
a neutrino flux which is an order of magnitude smaller than the one required
to explain the IceCube excess. This discrepancy could be explained by the 
slower diffusion of low-energy CRs which have not yet reached the GMC.

Limits
on the fraction of photons in the CR flux as e.g.\ those of CASA-MIA,
KASCADE and IceCube can be  used to constrain Galactic
neutrino sources~\cite{Ahlers:2013xia}. However, PeV $\gamma$-rays can
be absorbed both in sources and during propagation by pair production
on star light and CMB photons. Moreover, these gamma-ray limits are 
biased towards the northern hemisphere. As a result, they do not
exclude the case that (a fraction of) the IceCube excess has a Galactic
origin.

\section{Summary}

We have calculated the neutrino yield from collisions of CR nuclei on gas 
using the event generator QGSJET-II. Our numerical results assuming 
power-law fluxes for the primary CR nuclei can be used in all applications
where the neutrino yield is dominated by the contribution of heavy nuclei.
In the case of Galactic CRs, we found that the neutrino flux is well
approximated by accounting only for the proton component in the CR flux. 
Since the proton intensity above $10^{15}$\,eV varies considerably in 
different parametrisations of the elemental composition of Galactic CRs, 
the resulting variations in the predicted  neutrino intensity offer
the possibility to distinguish between these options.
The helium contribution in the interstellar medium to the neutrino flux
can be accounted for approximately by employing an enhancement factor, 
$\tilde\eps_{\rm M}\simeq 1.3$ around $E_\nu\sim {\rm PeV}$.

We have compared  the spectral shape and the magnitude of the predicted 
neutrino intensity to the IceCube excess. The slope of the 
neutrino intensity from interactions of sea CRs is close to
$\alpha=4.7$ both in the poly-gonato and the Hillas  model
at high energies, while it reflects the proton slope $\alpha=2.7$ at
low energies. The break is less pronounced in the escape model, 
where the neutrino intensity scales as $E^{-3.2}$ in the knee region.
Thus the expected slopes in the poly-gonato and the Hillas
model are  compatible with a cutoff (or break) in the energy spectrum 
of the IceCube events. However, the energy scale of the break,
the total number of events expected and their arrival directions make
an explanation of these events by Galactic sea CRs very unlikely.
By contrast, the energy spectrum of neutrinos produced by proton--gas 
interactions close to sources has the same slope as the CR injection 
spectrum. Since the  injection spectrum is flatter, this results in a
better agreement with the energy spectrum of the IceCube excess.

The required magnitude of the neutrino flux can be achieved,
if the sources illuminate a close GMC. Moreover, the sources have to be
relatively nearby, $d\lsim 1$\,kpc. A small distance to the sources 
would also explain, why the IceCube neutrino 
events are not as concentrated towards the Galactic plane as
expected. In conclusion, we consider it as a viable option that
the IceCube excess is {\em partly\/} connected to nearby Galactic CR sources.

Finally, let us comment on the position of the neutrino knee:
While its exact shape is model dependent, its position at 
$E_\nu\sim 10^{14}$\,eV reflects simply the fact that the maximal
neutrino energy in $pp$ interactions is approximately 10\% of
the energy of the proton primary. Our results for the Milky Way  
can be applied also to the neutrino energy spectrum expected from
other normal galaxies, in particular from starburst galaxies.
These galaxies have magnetic fields which are two orders of magnitude higher
than the Galactic magnetic field~\cite{beck}. As a consequence,
in the model of Ref.~\cite{GKS14} where the knee
is caused by the escape of CRs, its position and thus the one
of the neutrino knee is shifted by up too two orders of magnitude.
In contrast,  such a shift is not expected in models of the Hillas
type, because there the maximal rigidity to which the dominant population of 
supernovae can accelerate CRs is determined not by the ambient magnetic 
field but by the magnetic field created by CR instabilities.

\acknowledgments

We would like to thank Tom Gaisser for communications about the Hillas model.
We are grateful to Gwenael Giacinti and Dima Semikoz for discussions
and collaboration on CR propagation on which the ``escape model'' 
from Ref.~\cite{GKS14,more} is based. MK would like to thank
Nordita for hospitality and support during the program
``News in Neutrino Physics.''
SO acknowledges support from NASA through the grants NNX13AC47G and 
NNX13A092G.


\end{document}